# Cosmic Gamma-ray Bursts Studies with Ioffe Institute Konus Experiments

**R. L. Aptekar[1], S. V. Golenetskii, D. D. Frederiks, E. P. Mazets and V. D. Palshin**
*Ioffe Physico-Technical Institute*
*St. Petersburg, 194021, Russian Federation*
*E-mail:* `aptekar@mail.ioffe.ru`

We present a short review of GRB studies performed for many years by Ioffe Institute experiments onboard a number of space missions. The first breakthrough in the studies of GRB was made possible by four Konus experiments carried out by the Ioffe Institute onboard the *Venera 11-14* deep space missions from 1978 to 1983. A new important stage of our research is associated with the joint Russian-American experiment with the Russian Konus scientific instrument onboard the U.S. *Wind* spacecraft which has been successfully operating since its launch in November 1994. The Konus-*Wind* experiment has made an impressive number of important GRB observations and other astrophysical discoveries, due to the advantages of its design and its interplanetary location. We also briefly discuss future GRB experiments of the Ioffe Institute.

[1] Speaker

## 1. The Konus experiments on board *Venera* missions

The nature of cosmic gamma-ray bursts (GRBs) as extremely explosive releases of electromagnetic energy has been the focus of astrophysicists' attention since soon after their discovery in 1967-1973 by U.S. *Vela* satellites [1]. One of the earliest confirmations of the discovery of this new phenomenon was provided by observations collected by Ioffe Institute using data from *Kosmos 461* satellite [2]. An important breakthrough in studies of GRBs was made possible by four Konus experiments carried out by the Ioffe Institute on board the *Venera 11, 12, 13 & 14* deep space missions in 1978-83. The sensor system of each Konus instrument consisted of six scintillation detectors with close to cosine angular sensitivity pattern arranged along cartesian axes of the spacecraft. It gave an opportunity to determine the direction towards GRBs source independently using single spacecraft data. The *Venera* interplanetary missions were launched in pairs and separated each from other by a distance measured by several tens million kilometers. Such large distance between missions provided a high accuracy for GRB source triangulation on the celestial hemisphere. Therefore the Konus experiments had two independent methods for GRB source localization: autonomous approach based on data from detectors with anisotropic angular sensitivity and triangulation method used large distances between *Venera* missions.

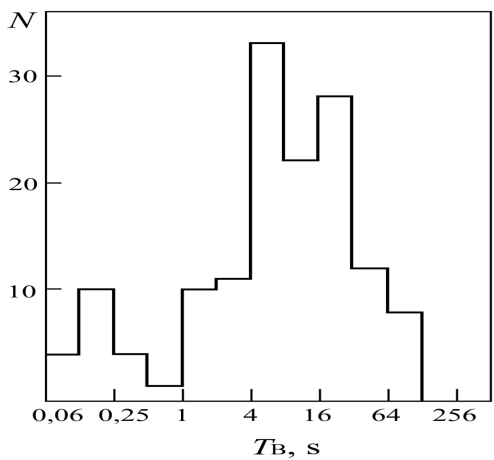
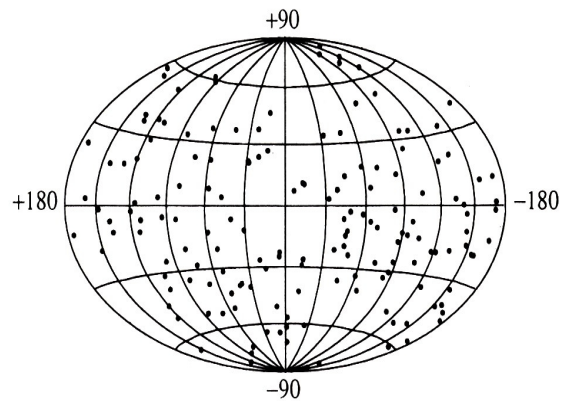

*Fig. 1. Konus observations of the GRBs temporal structures revealed the existence of a separate class of short bursts [3].*

*Fig. 2. Localizations of GRB sources demonstrated their isotropic distribution on the celestial sphere [4].*

The Konus experiments were among the first to make measurements of the various observational properties of GRBs, which still constitute the basis of modern concept in GRB research. The Konus observations of the temporal structures of gamma-ray bursts were central to the discovery of the existence of a separate class of short bursts, demonstrating the so-called "bimodal" duration distribution [3] (Fig. 1). For the first time with good statistical accuracy, it had been shown that distribution of GRB sources over the celestial sphere is random [4] (Fig. 2). Later, this result was confirmed with an even larger number of events in the well-known BATSE experiment onboard NASA's *Compton Gamma-Ray Observatory* [5]. Other important result of the Konus experiment was a demonstration of a strong correlation between temporal and spectral behavior of GRBs [6].

An extremely interesting and entirely unexpected result was that a new class of gamma-ray transients was discovered by the Konus experiment onboard *Venera 11-14* missions. The most peculiar was the burst of March 5, 1979 [7] (Fig. 3). It started with a very intensive initial burst with a hard radiation component. Then, soft pulsed radiation was observed. A prominent peculiarity of this event was a series of short repeating soft bursts observed in the Konus experiment. A total of 16 repeating bursts were recorded during Konus observations [8]. Later such sources have been named the soft gamma-ray repeaters (SGRs). The source of March 5 event obtained SGR 0526-66 notation. One else source of repeating bursts was discovered also on March, 1979 during the Konus experiment. It was SGR 1900+14 [9].

## 2. The Joint Russian-American Konus-*Wind* experiment

A new and important chapter in the GRB and SGR research at Ioffe Institute is associated with a joint Russian-American experiment [10] conducted with the Russian Konus instrument on board the U.S. *Wind* spacecraft. The deep space trajectory of the *Wind* spacecraft is exceedingly favorable for studies of GRBs and SGRs. Two high-sensitive scintillation detectors of the Konus-*Wind* gamma-spectrometer permanently view the entire celestial sphere, in such a way that no event important to the astrophysics of gamma-ray bursts and gamma-repeaters has yet been missed by Konus-*Wind* during all its successful and uninterrupted operation for more than 17 years. The instrument has an optimal program for detecting temporal and spectral characteristics of GRBs. The low energy threshold of the instrument is shifted for many years from 12 keV up 20 keV. The energy range of the instrument is now from 20 keV up to 15 MeV. Since the beginning of the experiment the Konus-*Wind* has detected 2198 GRBs in a triggered mode, 2104 GRBs in a waiting mode, 249 bursts from SGRs and 827 solar flares.

The Konus-*Wind* experiment takes part in simultaneous observations of GRBs with *Swift*/BAT, *Fermi*/GBM, *Suzaku*-WAM and other space GRB missions. Such observations give us an opportunity to perform joint spectral fits in a wide energy range. Such joint spectral fits were performed for a number of GRBs [11,12].

Owing to its high omnidirectional sensitivity and location far from the Earth, providing the optimum situation for observations of the entire sky at all times, the Konus-*Wind* experiment is an unique source of information about the temporal and spectral characteristics of gamma-ray events in the energy range from 20 keV to 15 MeV. These data constitute essential elements of modern multi-wavelength studies of gamma-ray transient events and include data obtained from spacecraft together with data from a network of ground-based optical and radio telescopes. Konus-*Wind* observations are thus in wide demand. There are two bright examples of such successful observations. The first is the observations of so-named "naked-eye" GRB 080319B [13]. Multi-wavelength studies of this burst demonstrate, that optical and gamma emissions begin and end at the same time, which strongly suggests that they originate from the same physical region. The optical and gamma emission of the GRB 050820A, obtained by RAPTOR telescope of Los-Alamos Laboratory and by Konus-*Wind* experiment, evidences that both type of radiation were detected simultaneously [14].

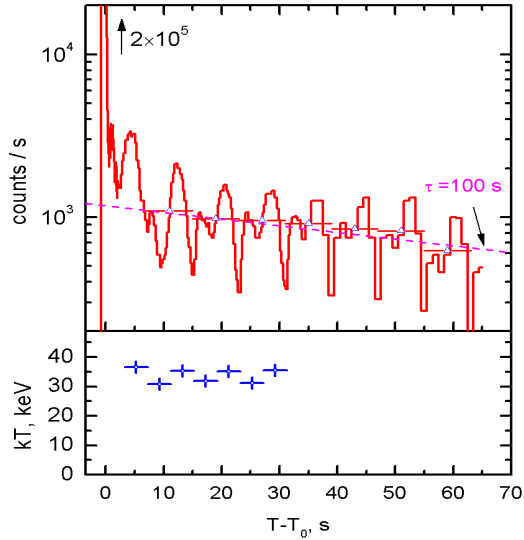 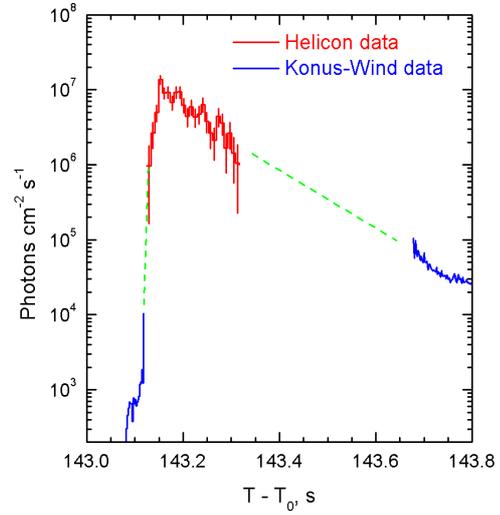

*Fig. 3. Giant flare from SGR 0526-66 was observed on March 5, 1979 [7].*

*Fig. 4. The temporal profile of the initial pulse of the giant flare from SGR1806-20.*

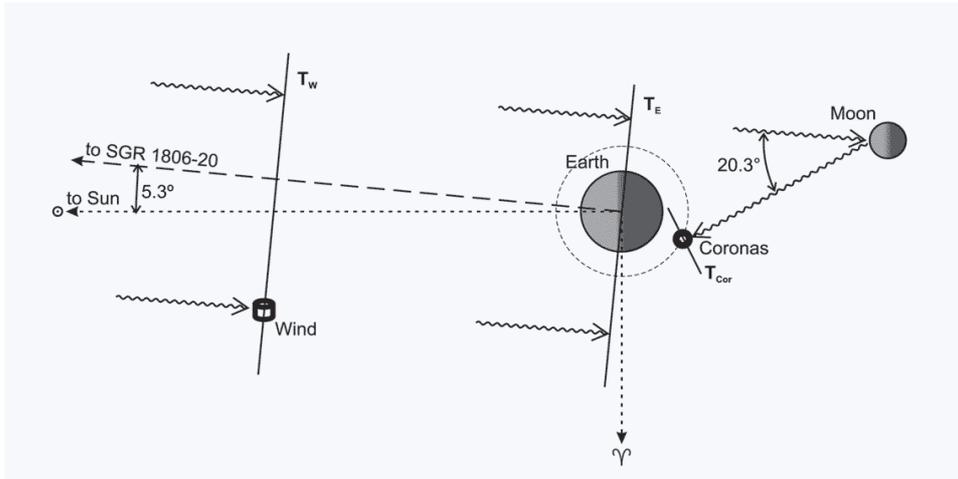

*Fig. 5. Reflection of the initial pulse of the giant flare from SGR1806-20 by the Moon [15].*

    A very important and unusual result was obtained during simultaneous observations of the giant flare from SGR 1806-20 on December 27, 2004 by Konus-*Wind* instrument and Helicon instrument on board Russian near-Earth solar observatory *CORONAS-F*. The photon counting rate of gamma-rays in the initial pulse of a giant SGR flare is always so great that the sensitive detectors of instruments are fully overloaded ("saturated") in such a way that precise measurements of the initial pulse become difficult and only rough lower-bound estimates are possible. Positions of the spacecraft at the time of detection of this burst are shown schematically in Fig 5. The detectors of the Helicon instrument were screened by Earth from direct exposure to the initial pulse of the giant flare from gamma-repeater, bur clearly recorded its reflection from the Moon surface. This reduction in intensity allowed, for the first time, reliably reconstructing the temporal profile of the initial pulse of the giant flare (Fig. 4) and

determining its energy parameters: the full isotropic energy release of $2.3 \times 10^{46}$ erg and the peak luminosity of $3.5 \times 10^{47}$ erg s$^{-1}$ [15]. The research on the December 27, 2004 giant flare became the first example of studying Moon-reflected X-ray and gamma-radiation coming from a source outside the Solar System. Recently, Konus-*Wind* measurements have been central in discovering SGRs outside our own Milky Way system in the nearby galaxies M81 and M31 [16,17].

One more recent result of the experiment is connected with GRB 110918A, which is the most intense long GRB in the history of Konus-*Wind* observations since November, 1994. The burst had a 20 keV – 10 MeV fluence of $7.5 \times 10^{-4}$ erg cm$^{-2}$, and 16 ms peak flux of $8.7 \times 10^{-4}$ erg cm$^{-2}$s$^{-1}$. GRB was localized by IPN and bright X-ray source was found on the edge of the IPN box in *Swift*/XRT TOO observations. A detailed analysis of the Konus-Wind prompt gamma-ray detection (Fig. 6), together with the refined IPN localization and results of the ~50 days-long afterglow monitoring by *Swift*/XRT and *Swift*/UVOT can be found soon in the forthcoming paper [18].

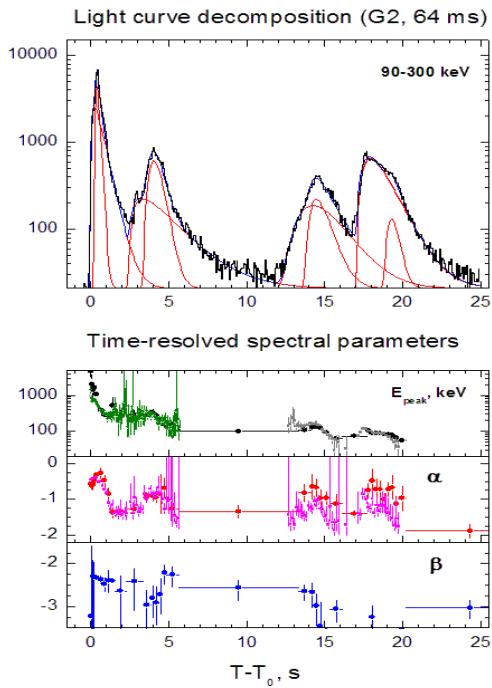
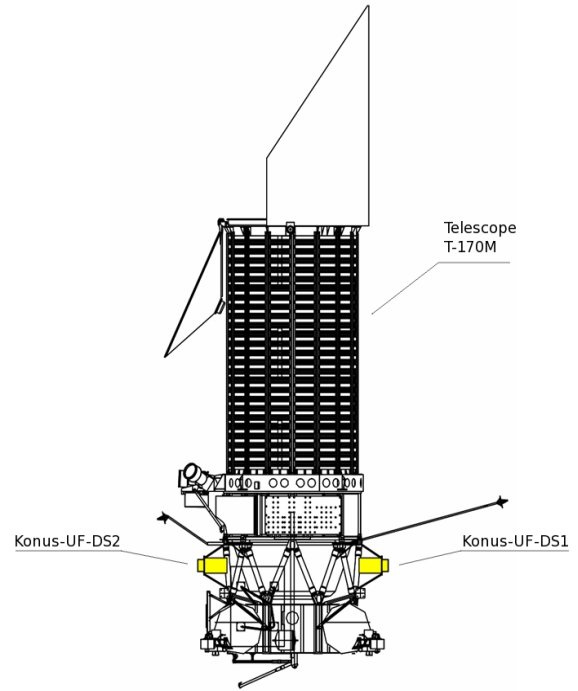

*Fig. 6. Light curve and time-resolved spectral parameters of GRB110918A.*

*Fig. 7. Allocation of Konus-UF detectors onboard the Spectr-UF mission.*

## 3. The future Ioffe Institute experiments in the field of GRB study

The Konus-UF is one of the future Ioffe Institute experiments aimed at comprehensive investigations of GRB prompt gamma-radiation. It is planned that the Konus-UF instrument will be accommodated on board Russian *Spektr-UF* mission which has the second name the World Space Observatory. The Konus-UF instrument consists of two detector units and an electronic unit. The Konus-UF detector units supplied by a NaI(Tl) scintillation crystal with dimensions like Konus-*Wind* detectors have. Each Konus-UF detector will be allocated on the spacecraft in such a manner as to observe the half of hemisphere (Fig. 7). The energy range of Konus-UF instrument is from 10 keV up to 15 MeV. The instrument is going to have a detailed program for measurement of time and spectral characteristics of GRBs. It is planned that the *Spektr-UF* mission will be launched in 2016 to a high-apogee orbit.

## 4. Conclusions

The early Konus experiments onboard *Venera 11-14* deep space missions had firstly revealed many of the basic characteristics of GRBs. The Joint Russian-American Konus-*Wind* experiment, which has already been operating for more than 17 years, provides important and often unique data regarding the various characteristics of GRBs in the 20 keV to 15 MeV energy range. The Konus-UF experiment is planned for launch in 2016. It will give us an opportunity to continue effective research into explosive phenomena in the Universe.

R.L. acknowledges the Organizing Committee of the Fermi/Swift GRB conference 2012 for the support of the participation in the conference.

**References**

[1] Klebesadel, R.W., Strong, I.B., & Olson, R.A., *Observations of Gamma-Ray Bursts of Cosmic Origin*, ApJ **182** (1973) L85.

[2] Mazets, E.P., Golenetskii, S.V., & Il'inskii, V.N., *Flare of cosmic gamma radiation as observed with "Cosmos-461" satellite*, JETP Lett. **19** (1974) 77.

[3] Mazets, E.P., et al., *Catalog of cosmic gamma-ray bursts from the KONUS experiment data*, Astrophysics and Space Science **80** (1981) 3.

[4] Mazets, E.P., & Golenetskii, S.V., *Observations of cosmic gamma-ray bursts*, Sov. Sci. Rev. Sect. E **6** (1988) 283.

[5] Paciesas, W.S., et al., *The Fourth BATSE Gamma-Ray Burst Catalog (Revised)*, ApJS. **122** (1999) 465.

[6] Golenetskii S.V. et al., *Correlation between luminosity and temperature in gamma-ray burst sources*, Nature **306** (1983) 451.

[7] Mazets E.P. et al., *Observations of a flaring X-ray pulsar in Dorado*, Nature **282** (1979) 587.

[8] Golenetskii S.V. et al., *Recurrent bursts in GBS0526 - 66, the source of the 5 March 1979 gamma-ray burst*, Nature **307** (1984) 41.

[9] Mazets, E.P., et al., *Soft gamma-ray bursts from the source B1900+14*, Sov. Astron. Lett. **5** (1979) 343.

[10] Aptekar, R.L., et al., *Konus-W Gamma-Ray Burst Experiment for the GGS Wind Spacecraft*, Space Science Reviews **71** (1995) 265.

[11] Krimm, H., et al., *GRB 050717: A Long, Short-Lag, High-Peak Energy Burst Observed by Swift and Konus*, ApJ **648** (2006) 1117.

[12] Ohno, M., et al., *Spectral Properties of Prompt Emission of Four Short Gamma-Ray Bursts Observed by the Suzaku-WAM and the Konus-Wind*, PASJ **60** (2008) 361.

[13] Racusin, J.L., et al., *Broadband observations of the naked-eye γ-ray burst GRB080319B*, *Nature* **455** (2008) 183.

[14] Vestrand, W.T., et al., *Energy input and response from prompt and early optical afterglow emission in γ-ray bursts*, *Nature* **442** (2006) 172.

[15] Frederiks, D.D., et al., *Giant flare in SGR 1806-20 and its Compton reflection from the Moon*, *Astron. Lett.* **33** (2007) 1.

[16] Frederiks, D.D., et al., *On the possibility of identifying the short hard burst GRB 051103 with a giant flare from a soft gamma repeater in the M81 group of galaxies*, *Astron. Lett.* **33** (2007) 19

[17] Mazets, E.P., et al., *A Giant Flare from a Soft Gamma Repeater in the Andromeda Galaxy (M31)*, *ApJ* **680** (2008) 545.

[18] Frederiks, D.D., et al., *The ultra-luminous GRB 110918A*, *in preparation* (2012).